\documentclass{desyproc}
\usepackage{graphicx}

\begin{document}
\title{Lepton-number violating meson decays\\ in theories beyond the Standard Model}

\author{{\slshape Anatoly Borisov}\\[1ex]
Faculty of Physics, Moscow State University, 119991 Moscow, Russia}

\contribID{borisov\_anatoly}

\desyproc{DESY-PROC-2009-07}
\acronym{HQP08} 

\maketitle

\begin{abstract}
After discussion of mechanisms of lepton number violation, we consider meson decays $K^+ \to\pi^ - \ell ^ + \ell '^ +$ and $D^ + \to K^ - \ell ^ + \ell '^ +$ ($\ell,\ell ' = e,\mu $) with $\Delta L= 2$ in the Standard Model extended by massive Majorana neutrinos and in a supersymmetric extension of the Standard Model with explicit breaking of $R$-parity by trilinear or bilinear Yukawa couplings in the superpotential.  We give estimates of the branching ratios for these decays and compare the effectiveness of various decay mechanisms taking into account present experimental bounds on lepton mixing, masses of neutrinos and superparticles, and $R$-parity violating couplings.
\end{abstract}

\section{Introduction}

In the Standard Model (SM), the lepton $L$ and baryon $B$ numbers are conserved due to the accidental $
U(1)_L  \times U(1)_B$ symmetry. But the $L$ and $B$ nonconservation is a generic feature of various extensions of the SM.
That is why  lepton-number violating processes are sensitive tools for testing  theories beyond the SM.

The following $\Delta L\neq 0$ processes have been extensively studied:
neutrinoless double beta decay $ (A,Z) \to (A,Z + 2) + e^ -   + e^ -$~\cite{Fur,EV,ABZh1};
rare decays of mesons ${ M^ +   \to M'^ -  \ell ^ +  \ell '^ +}$ $\left( {\ell ,\ell ' = e,\mu } \right)$ ($ K^+
\to \pi^-\mu^+\mu^+$ etc.)~\cite{LS1,LS2,ABZ1,ABS1,ABS2,ABS3} and baryons $\Xi^-\to p\,\mu^-\ell^-$, $\Sigma^-\to p\,\mu^-\ell^-$ $(\ell =e,\mu)$, $\Xi^+_c\to \Xi^-\mu^-\mu^-$ etc.~\cite{LS3};
same-sign dilepton production in high-energy hadron-hadron and lepton-hadron collisions:
$ pp \to \ell ^ \pm  \ell '^ \pm X$~\cite{ABZ2,Pan,CGZh}, $ e^\pm p\to \mathop {\nu _e }\limits^{( - )}\ell ^ \pm  \ell '^ \pm X$~\cite{Rod,ABZh};
$ (\mu^-,e^+)$ conversion in nuclei $ (A,Z) + \mu_b^- \to e^{+} + (A,Z-2)^{*}$~\cite{DKFS}.

The SM is based on the gauge group
\begin{equation}
G_{\rm SM} = {SU(3)_c}\times {SU(2)_I}\times  {U(1)_Y}
\label{GSM}
\end{equation}
with the subscripts $c$, $I$, and $Y$ denoting color, weak isospin and hypercharge, respectively, and three fermion generations (families, $f=1,2,3$), each of them is consisted of 5 different representations of $G_{\rm SM}$:
\begin{eqnarray}
&L^f_L( {\bf 1}, {\bf 2}, { -1}) = (\nu_{eL}, e_L)^T, (\nu_{\mu L}, \mu_L)^T, (\nu_{\tau L}, \tau_L)^T;\nonumber\\
&Q^f_L( {\bf 3}, {\bf 2}, 1/3) = (u_L, d_L)^T, (c_L, s_L)^T, (t_L, b_L)^T;\nonumber\\
&E^f_R( {\bf 1}, {\bf 1}, { -2}) = e_R, \mu_R, \tau_R;~
U^f_R( {\bf 3}, {\bf 1}, 4/3) = u_R, c_R, t_R;~
D^f_R( {\bf 3}, {\bf 1}, {-2/3}) = d_R, s_R, b_R.
\label{SMc}
\end{eqnarray}

The fermion interactions are mediated by $12~(= 8_c + 3_I + 1_Y)$ gauge vector bosons.

In addition, the SM contains a {\it single} Higgs boson doublet $\varphi ({\bf 1},{\bf 2},{1})$. Its {\it nonzero} vacuum expectation value $\left\langle \varphi  \right\rangle$ {\it spontaneously breaks} the gauge symmetry and {\it yields masses} to weak bosons (as well as to charged fermions and the Higgs boson itself):
\begin{equation}
\left\langle \varphi  \right\rangle  = (0, v/\sqrt 2~)^{T} \Rightarrow \quad G_{\rm SM} \to  { SU(3)_c}\times   U(1)_{Q},
\label{SSB}
\end{equation}
where the electric charge $Q =  I_3  + Y/2$.

{\it The Higgs boson is the only piece of the SM which is not confirmed experimentally up to now!}

\section{Lepton numbers and mechanisms of their violation}

So there are  three lepton families (generations) in the SM (see Eq. (\ref{SMc})). By definition, the {\it lepton family number} (LFN) $L_\ell = + 1 (-1)$ for particles $\ell = e^-, \nu_e,  \ldots$ (for antiparticles ${\bar\ell} = e^+, {\bar\nu}_e,  \ldots$), and $L_\ell =0$ for leptons $\ell'\neq \ell$, ${\bar\ell}'\neq {\bar\ell}$.
The {\it total lepton number} (LN)
\begin{equation}
L=L_e + L_\mu + L_\tau,
\label{LN}
\end{equation}
so that $L=+1 (-1)$ for each $\ell$ (${\bar\ell}$) and $L= 0$ for other particles ({\it non}leptons).

In the minimal SM (with {\it massless} neutrinos), each LFN is {\it conserved  separately}. For example, in the muon decay $\mu ^ -\to e^ -   + \bar \nu _e   +  \nu _\mu$: $L_e = 0  =  1  + ( - 1) +  0$,~  $L_\mu  =  1   =  0   +   0   +   1$,  ~
$L=  1  = 1  +  ( - 1) +1$.

The SM has three {\it active} neutrinos $\nu _{\ell L} (\ell =e,\mu,\tau)$ taking part in charged current (CC) and neutral current (NC) weak interactions mediated by the massive charged $W^ \pm$ and neutral $Z$ bosons:
\begin{equation}
{\cal L}_{\rm CC}  =  - \frac{g}{{\sqrt 2 }}\sum\nolimits_\ell  {\left( {\bar \ell _L \gamma ^\mu  \nu _{\ell L} W_\mu ^ -   + \bar \nu _{\ell L} \gamma ^\mu  \ell _L W_\mu ^ +  } \right)} ,~
{\cal L}_{\rm NC}  =  - \frac{g}{{2\cos \theta _W }}\sum\nolimits_\ell  {\bar \nu _{\ell L} \gamma ^\mu  \nu _{\ell L} Z_\mu},
\label{CN}
\end{equation}
where the weak-mixing angle is defined by $\tan \theta _W = g'/g$ with $g$ and $g'$ being the the $SU(2)_I$ and $U(1)_Y$ gauge couplings, respectively.
The SM contains no {\it sterile} neutrinos $\nu _{\ell R}$.

In the SM, the lepton family $L_\ell$ and baryon $B$ numbers are conserved to all orders of perturbation theory due to the {\it accidental} global symmetry:
\begin{equation}
G_{\rm SM}^{\rm global} =U(1)_{L_e}\times U(1)_{L_\mu}\times U(1)_{L_\tau}\times U(1)_{B},
\label{GSMg}
\end{equation}
existing at the level of {\it renormalizable operators}. The symmetry (\ref{GSMg}) is called {\it accidental} because we  do not impose it {\it intentionally}. It is a {\it direct consequence} of the gauge symmetry and the choice of the representations of the physical fields.

The SM is a {\it chiral} gauge theory, since there are $L$-doublets and $R$-singlets of the gauge group $SU(2)_{I}$ (they have {\it different} electroweak interactions, see Eqs. (\ref{SMc}) and (\ref{CN})).

The {left-handed} and {right-handed} chiral components of a Dirac field $\psi$ are defined as:
\[
\psi _{L} = P_{L}\psi,\quad \psi _{R} = P_{R}\psi,\quad \psi = \psi _{L} + \psi _{R},
\]
where the chirality projection operators
\[
P_{L,R} = (1 \mp \gamma ^5)/2 = P_{{L},{R}}^2,\quad P_{L}P_{R} =0,\quad \gamma ^5  = i\gamma ^0 \gamma ^1 \gamma ^2 \gamma ^3.
\]

{\it Chirality} is eigenvalue of  the operator  $\gamma ^5$: $ \gamma ^5 \psi _{L}  =  - \psi _{L},\quad \gamma ^5 \psi _{R}  =  + \psi _{R}$.

Taking into account the relations:
\begin{equation*}
\bar \psi \gamma ^\mu  \partial _\mu  \psi  = \bar \psi _{L} \gamma ^\mu  \partial _\mu  \psi _{L}  + \bar \psi _{R} \gamma ^\mu  \partial _\mu  \psi _{R};~\bar \psi \psi  = \bar \psi _{R} \psi _{L}  + \bar \psi _{L} \psi _{R},
\end{equation*}
where $\bar \psi _{{L}, {R}}  = \overline{P_{{L},{R}}\psi}  = {\bar \psi}P_{{R},{L}},~\bar \psi  = \psi ^ +  \gamma ^0$, we see:

$\bullet$ chiral components interact with gauge fields {\it independently};

$\bullet$ the Dirac mass term (${\cal L}_D = -m_D {\bar\psi}\psi$) in the Lagrangian relates {\it different} chiral components and {\it violates} chirality conservation that takes place for {\it massless} (Weyl) fermions.

In the SM, neutrinos are massless due to absence of $\nu_{\ell R}$. The only possible neutrino mass term ${\cal L}_{ML}  =  - \frac{1}{2}m_L (\bar \nu _L \nu _L^c  + \bar \nu _L^c \nu _L )$ ~ {\it violates} the lepton number: $\Delta L = \pm 2$. The global symmetry (\ref{GSMg}) prevents generation of the Majorana mass term ${\cal L}_{ML}$ by loop corrections.

The $B-L$-violating terms {\it cannot be induced even  nonperturbatively} because the $U(1)_{B-L}$ subgroup of the group  (\ref {GSMg}) is {\it non-anomalous}.

Discovery of neutrino oscillations (1998--2002) (predicted by B. Pontecorvo in 1957 \cite{BP}),
\[
\nu_{\ell} \to \nu_{ \ell'}\, ({\ell} \neq {\ell'}),
\]
has clearly demonstrated the LFN violation: $\Delta L_{\ell'}  =  - \Delta L_{\ell}  = 1$.
Here $\nu_{\ell}$ is the neutrino of flavor $\ell  = e,\mu ,\tau $. It is created in association with the charged lepton  ${\ell ^ +  }$
in the decay $W^ +   \to {\ell ^ +}   + \nu _{\ell}$.

Up to now the oscillations have been observed unambiguously for solar ($\nu _e  \to \nu _\mu  (\nu _\tau  )$), atmospheric ($\nu _e  \to \nu _\mu  (\nu _\tau  )$), reactor ($\bar \nu _e  \to \bar \nu _\mu  $), and accelerator ($\nu _\mu   \to \nu _\tau  $) neutrinos (for a review, see \cite{PDG-08}).

The neutrino oscillations imply that {\it neutrinos are massive and mixed particles}, i.e. the neutrino  flavor state is a  coherent superposition of neutrino mass eigenstates:
\begin{equation}
\left|\nu _{\ell}\right\rangle   = \sum\nolimits_i {U_{{\ell} i}^{*}\left|\nu _i\right\rangle} \quad ({ \ell  = e,\mu ,\tau }).
\label{fs}
\end{equation}
Here $U = (U_{{\ell} i})\equiv U_{\ PMNS}$ is the Pontecorvo--Maki--Nakagawa--Sakata lepton mixing matrix \cite{BP,PMNS}, $\nu _i$s are neutrinos with  definite masses $m_i$, and the neutrino mass spectrum is {\it nontrivial}: $\Delta m_{jk}^2  \equiv m_j^2  - m_k^2 \neq 0$.

So neutrino oscillations {\it require extension} of the SM (New Physics) to include nonzero neutrino masses and violation of LFNs.
One of the main unsolved questions of particle physics is the nature of neutrino masses: to be Dirac or Majorana type? It should be noted that the neutrino oscillations do not probe the nature of the mass.

The Dirac neutrino carries the lepton number which distinguishes it from the antineutrino.
The  Dirac neutrino mass term ${\cal L}_D$ is generated just like the quark and charged lepton masses via the standard Higgs mechanism (see Eq. (\ref{SSB})) with addition of right-handed neutrinos $\nu_{\ell R}$:
\begin{eqnarray}
& - {\cal L}_{\rm Yuk}  = y_{\ell\ell'} \bar L_\ell \tilde \varphi \nu _{\ell'R}  + {\rm H.c.}, ~
\bar L_\ell = (\bar \ell _{\ell L} ,\bar \nu _{\ell L} ),\quad \tilde \varphi  = i\tau _2 \varphi  \Rightarrow \tilde \varphi _0  =
(v/\sqrt 2, 0)^{T},\nonumber\\
& - {\cal L}_{Yuk}  \Rightarrow  - {\cal L}_D  = \left( {M_D } \right)_{\ell\ell'} \bar \nu _{\ell L} \nu _{\ell' R}  + \mbox{H.c.},
\label{MD}
\end{eqnarray}
where $y_{\ell\ell'}$ are Yukawa couplings. The Dirac mass matrix is complex and {\it nondiagonal}:
$\left( {M_D } \right)_{\ell\ell'}  = y_{\ell\ell'}v/\sqrt{2}$.
Therefore ${\cal L}_D$ {\it violates} LFNs $L_e, L_\mu, L_\tau$, but it {\it conserves} the total LN (\ref{LN}).
The standard diagonalization gives ${\cal L}_D  =  - \sum\nolimits_{i}{m_i \bar \nu _i } \nu _i$,
where $\nu _i$ is the 4-component field of Dirac neutrinos with mass $m_i$, and flavor fields in Eqs. (\ref{CN})
\[
\nu _{\ell L} (x) = \sum\nolimits_{i}{U_{\ell i} } \nu _{iL} (x),
\]
$U$ is the PMNS mixing matrix (see Eq. (\ref{fs})).

The Majorana neutrino is a true neutral particle identical to its antiparticle \cite{Maj}.
There are two types of  Majorana mass terms (we consider a simple case of one flavor):
\begin{equation}
{\cal L}_{ML}  =  - \frac{1}{2}m_L (\bar \nu _L^c \nu _L^{}  + \bar \nu _L^{} \nu _L^c ),\;\quad {\cal L}_{MR}  =  - \frac{1}{2}m_R (\bar \nu _R^c \nu _R^{}  + \bar \nu _R^{} \nu _R^c ).
\label{MLR}
\end{equation}
Here the {\it charge conjugated} fields are defined as follows:
\[
\psi ^c  = C\bar \psi ^T  = C\gamma ^{0T} \psi ^ *  (\psi ^ *= (\psi^+)^T) , \quad \bar \psi ^c \equiv \overline{\psi ^c} = \psi^{T}C = - \psi^{T}C^{-1},\quad C = i\gamma ^2 \gamma ^0 ,
\]
and useful relations are valid: $ \psi_{L}^{c} \equiv \left( {\psi_{L} } \right)^{c}  = P_{R}\psi ^{c}  = \left( {\psi ^{c} } \right)_{R},~
\psi_{R}^{c}  \equiv \left( {\psi _{R} } \right)^{c}  = P_{L}\psi ^{c}  = \left( {\psi ^{c} } \right)_{L}$.

The Majorana mass term {\it violates lepton number by two units}, $\Delta L =  \pm 2$.

The total Dirac--Majorana mass term is given by	(see Eqs. (\ref{MD}) and (\ref{MLR}))
\begin{equation}
{\cal L}_{D + M}  = {\cal L}_{D} + {\cal L}_{ML} + {\cal L}_{MR},
\label{D+M}
\end{equation}
and after diagonalization it takes the form
\[
 {\cal L}_{D + M}  =  - \frac{1}{2}\sum\nolimits_k {m_k^{} } (\bar \nu _{kL}^c \nu _{kL}^{}  + \bar \nu _{kL}^{} \nu _{kL}^c)  =  - \frac{1}{2}\sum\nolimits_k {m_k^{} } \bar \nu _k^{} \nu _k^{} ,
 \]
where two mass eigenstates, $\nu _k^{}  = \nu _{kL}^{}  + \nu _{kL}^c  = \nu _k^c$, are Majorana neutrinos.

From experimental data, we know that  the masses of observed neutrinos are much smaller than those of charged leptons ($m_\ell$) and quarks ($m_q$): $0.04~{\rm eV} < {\rm Mass}~[{\rm Heaviest}~\nu_i] < (0.07 \div 0.7)~{\rm eV}$ \cite{PDG-08}. The dominant paradigm for the origin of {\it finite but tiny} neutrino mass is the {\it seesaw mechanism} (for a review, see \cite{GK}): beyond the SM (at ultra-high energies) there exists a mechanism generating the {\it right-handed Majorana mass term}, and the Dirac mass term is generated through the standard Higgs mechanism, so that in Eq. (\ref{D+M})
\begin{equation}
m_R  \gg m_D \sim m_\ell~ {\rm or}~m_q,~m_L = 0.
\label{seesaw}
\end{equation}
The neutrino $\nu _R $  is {\it completely neutral} under the SM gauge group (\ref{GSM}), and  $m_R $ is {\it not connected} with the SM symmetry breaking scale $v = \left( {\sqrt{2} G_F } \right)^{ - 1/2}  \simeq 246~{\rm{GeV}}$, but is associated to a different higher mass scale, e.g., the GUT-scale: $m_R  \sim \Lambda_{\rm GUT} \sim 10^{15}  \div 10^{16}~{\rm{GeV}} \gg m_D$. There exists a large number of seesaw models in which both $m_D $  and  $m_R $  vary over many orders of magnitude, with $m_R $ ranging somewhere between the TeV scale and the GUT-scale \cite{Lan}.

Diagonalization of the mass term (\ref{D+M}) of the type (\ref{seesaw}) gives two mass eigenstates, which are light $\nu_1$ and heavy $\nu_2$ Majorana neutrinos:
\begin{eqnarray*}
& m_1  \simeq m_D^2/m_R \ll m_D ,\quad m_2  \simeq m_R  \gg m_D ;\\
& \nu _L  = i\nu _{1L} \cos \theta  + \nu _{2L} \sin \theta ,~
 \nu _R^c  =  - i\nu _{1L} \sin \theta  + \nu _{2L} \cos \theta ,~
 \tan 2\theta  = 2m_D /m_R  \ll 1,
\end{eqnarray*}
so that $\nu _L  \simeq i\nu _{1L},~ \nu _R^c \simeq \nu _{2L}$.

In the general case of an {\it arbitrary} number $n_s (\geq 3)$ of {\it electroweak-singlet (sterile)} neutrinos, the seesaw mass term is
\begin{equation}
-{\cal L}_{{D} + {MR}}  =  \bar \nu _L {M_D} \nu _R  + \frac{1}{2}\bar \nu _R^c {\ M_R} \nu _R  + {\rm H.c.},
\label{DMR}
\end{equation}
where $M_D $   is a {$3 \times n_s $ Dirac mass matrix} and  $M_R $ is a  {$n_s  \times n_s $ Majorana mass matrix}.
Its diagonalization  by means of a unitary  $\left( {3 + n_s } \right) \times \left( {3 + n_s } \right)$  matrix $V$ gives 3 light and $n_s $ heavy Majorana neutrinos:
\begin{equation}
\nu _{\ell L}  = \sum\nolimits_{k = 1}^3 {V_{\ell k} \nu _{kL}^{{\rm{light}}} }  + \sum\nolimits_{k = 4}^{n_s  + 3} {V_{\ell k} } \nu _{kL}^{{\rm{ heavy}}}.
\label{lh}
\end{equation}

A possible scenario of the generation of the Dirac-Majorana mass term ${\cal L}_{D + MR}$ suitable for the seesaw mechanism may look as follows:
the grand unified group $G_{\rm GUT} = SO(10)$  can be {\it broken} to the SM group $G_{\rm SM}$ (\ref{GSM}) through the chain
\[
SO(10)\xrightarrow{\Lambda _{\rm GUT}} G_{\rm SM}  \times U(1)_{B - L} \xrightarrow{ V} G_{\rm SM} \xrightarrow{v}{SU\left( 3 \right)_c}  \times {U(1)_{Q}},
\]
with the {\it breaking scales} $\Lambda _{\rm GUT}$, $V (\sim 1 \div 10~{\rm TeV})$ and $v$. The generated mass matrices in Eq. (\ref{DMR}) are $M_R  = YV/\sqrt{2}$ and $M_D  = yv/\sqrt{2}$, where $Y$ and $y$ are the matrices of corresponding Yukawa couplings.

Probable mechanisms of LN violation may include exchange by:

$\bullet$ Majorana neutrinos \cite{GK} (the preferred mechanism after the discovery of neutrino oscillations \cite{PDG-08}: SM + $\nu_M$);

$\bullet$ SUSY particles (RPV MSSM \cite{Bar}: neutralinos, sleptons, squarks, gluinos);

$\bullet$ scalar bilinears \cite{BL} (the component $\xi^{--}$ of the $SU(2)_I$ triplet Higgs scalar, doubly charged dileptons etc.);

$\bullet$ leptoquarks \cite{LQ} (in various extensions of the SM:  scalar or vector particles carrying both $L$ and $B$ numbers);

$\bullet$ right-handed $W_R$ bosons in the {\it left-right symmetric} models \cite{LR} based on the gauge group $G_{LR} = SU(3)_c\times SU(2)_L\times SU(2)_R\times U(1)_{B-L}$ ($ \to G_{\rm SM} \to SU(3)_c \times U(1)_Q)$,  $\nu_R$'s and  the seesaw mechanism are needed);

$\bullet$ other (Kaluza--Klein sterile singlet neutrinos in  theories with large extra dimensions \cite{KK}: an  infinite tower of KK neutrino mass eigenstates, ...).

\section{Semileptonic Decays of pseudoscalar mesons\\ with $\Delta L = 2$}

As  examples of the processes with LN violation we consider the rare meson decays
\begin{equation}
K^+\to \pi^{\prime -} \ell^+ \ell^{\prime +},\quad  D^+\to K^{\prime -} \ell^+ \ell^{\prime +}\quad (\ell,~ \ell^\prime = e,\mu,\tau)
\label{dec}
\vspace{-0.2cm}
\end{equation}
mediated by Majorana neutrinos or supersymmetric particles.

\subsection{Decays via exchange by Majorana neutrinos}
The lowest order amplitude of the process is given by the sum of the tree  and
box diagrams shown in Fig.~1 (there are also two crossed diagrams with interchanged lepton lines).
\begin{figure}[htbp]
\vspace{-0.2cm}
\centering
\includegraphics[scale=0.7]{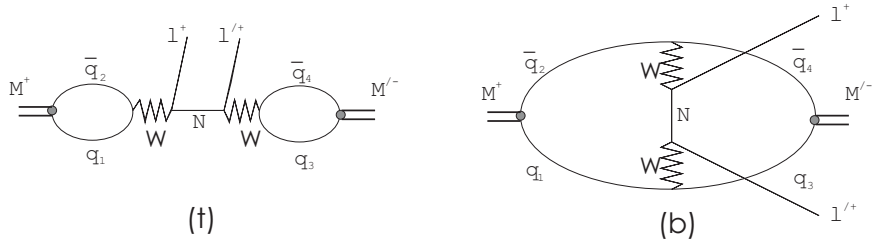}
\vspace{-0.3cm}
\caption{Feynman diagrams for the decay $ M^+\to
M^{\prime -} \ell^+ \ell^{\prime +}$. Here $N$ is a Majorana
neutrino, bold vertices correspond to Bethe--Salpeter amplitudes
for mesons as bound states of a quark and an antiquark.}
\label{Fig:decM}
\vspace{-1.0cm}
\end{figure}
\newpage
The width of the decay $M^{+}(P)\to M ^{\prime
-}(P^\prime)\ell ^{+}(p)\ell ^{\prime +}(p^\prime)$ is given by
\[
\Gamma _{\ell \ell '}  = \left( {1 - \frac{1}
{2}\delta _{\ell \ell '} } \right)\int {\left( {2\pi } \right)} ^4 \delta ^{(4)} (P' + p + p' - P)\frac{{\left| {A_t  + A_b } \right|^2 }}
{{2m_M }}\frac{{d^3 P'd^3 pd^3 p'}}
{{2^3 (2\pi )^9 P'^0 p^0 p'^0 }},
\]
where  $A_t$ ($A_b$) is the tree(box)-diagram amplitude expressed in the Bethe--Salpeter formalism of Ref.
\cite{Est} as
\[
A_i  = (2\pi )^{-8}\int {d^4 qd^4 q'{H_{\mu \nu }^{(i)} }{L_i^{\mu \nu} }} \quad (i = t,b).
\]
Here the lepton tensors
\begin{eqnarray}
&\displaystyle \ L_i^{\mu \nu }  = \frac{{g^4 }}
{4}\frac{{g^{\mu \alpha } }}
{{p_i^2  - m_W^2 }}\frac{{g^{\nu \beta } }}
{{p_i^{\prime 2}  - m_W^2 }}\sum\nolimits_N {U_{\ell N} } U_{\ell' N} \eta _N m_N \nonumber \\
&\displaystyle   \times \left( {\bar v^c (p)\left[ {\frac{{\gamma _\alpha  \gamma _\beta  }}
{{(p_i  - p)^2  - m_N^2 }} + \frac{{\gamma _\beta  \gamma _\alpha  }}
{{(p_i  - p')^2  - m_N^2 }}} \right]P_L v(p')} \right) \quad (i = {t}, {b}),
\label{Li}
\end{eqnarray}
where $\eta _N $ is  the  charge conjugation phase factor of the field of Majorana neutrinos with mass $m_N$:
$N  = \eta _N N^c ,\;\left| {\eta _N } \right| = 1$; $p_{ t}  = P,\;p'_{t}  = P'; p_{b}  = \frac{1}
{2}(P - P') + {q'} - {q},\;p'_{b}  = \frac{1}{2}(P - P') - {q'} + {q}$;
the hadron tensors
\begin{equation*}
H_{\mu \nu }^{({t})}  = {\rm Tr}\left[{\chi _P (q)}V_{12} \gamma _\mu  P_L \right]{\rm Tr}\left[ {{\bar \chi} _{P'} (q')}V_{43} \gamma _\nu  P_L\right],~
H_{\mu \nu }^{({b})}  = {\rm Tr}\left[ {\chi _P (q)}V_{13} \gamma _\mu P_L{{\bar \chi} _{P'} (q')}V_{42} \gamma _\nu  P_L \right]
\end{equation*}
are expressed in terms of the elements $V_{jk}$ of the CKM matrix and the model-dependent Bethe--Salpeter (BS) amplitudes
for the mesons \cite{Est}
\begin{equation}
\chi _P (q) = \int {d^4 xe^{iq \cdot x} } \chi _P (x) = \gamma ^5 (1 - \delta _M \hat P)\phi _P (q),
\label{BS}
\end{equation}
where $\delta _M  = (m_1  + m_2 )/m_M^2 $, $m_M$ is the mass of a meson made of a quark $q_1$ and an antiquark  $\bar q_2$ with {\it current} masses $m_1$ and  $m_2$, $q = (p_1  - p_2 )/2$ is the relative 4-momentum, $P = p_1  + p_2$ is the total 4-momentum of the meson, $\hat P = \gamma ^\mu  P_\mu$;  the function $\phi _P (q)$ is {\it model dependent}.
The tree amplitude is expressed in a {\it model independent} way in terms of the decay constants of the initial and final meson,  $f_M$  and $f_{M'}$, as follows:
\[
{A_t}  =  - \frac{1}
{4}{f_M f_{M'}} V_{12} V_{43} P_\mu  P'_\nu  {L_t^{\mu \nu }} .
\]
The box amplitude depends (in general) on the {\it details of hadron dynamics}
\begin{eqnarray*}
& A_b  = 2V_{13} V_{42} \delta _M \delta _{M'} (P_\mu  P'_\nu   + P_\nu  P'_\mu   - g_{\mu \nu } P \cdot P' + i\varepsilon _{\mu \nu \alpha \beta } P^\alpha  P'^\beta  )\\
&\displaystyle \times \int {\frac{{d^4 {q}}}
{{(2\pi )^4 }}} \frac{{d^4 {q'}}}
{{(2\pi )^4 }}{\phi _P (q)\phi _{P'} (q')}L_b^{\mu \nu } (q - q',p,p';P - P').
\end{eqnarray*}
We use the leading current-current approximation in the lepton tensors (\ref{Li}) due to relative smallness of the meson masses,  $m_M  \ll m_W $,
and the expression of the meson decay constant $f_M$ through the function $\phi _P (q)$ in the BS amplitude (\ref{BS}):
\[
f_M  = 4\sqrt {N_c }\, \delta _M \int {\frac{{d^4 q}}
{{(2\pi )^4 }}\phi _P (q)},
\]
where $N_c = 3$ is the number of quark colors. For the function $\phi _P (q)$, the relativistic Gaussian model has been used \cite{Est}:
\begin{eqnarray}
&\displaystyle \phi _P (q) = \frac{{4\pi }}
{{\alpha ^2 }}\left( {1 - {\mu} ^2 } \right)^{ - 1/2} \exp \left\{ { - \frac{1}
{{2{ \alpha ^2} }}\left[ {2\left( {\frac{{P \cdot q}}
{{m_M }}} \right)^2  - q^2 } \right]} \right\},\nonumber\\
&\displaystyle  { \alpha ^2}  = \frac{\pi }
{{4\sqrt {N_c } }}\left( {1 - {\mu} ^2 } \right)^{1/2} \frac{{f_M }}
{{\delta _M }},\;\;{ \mu}  = m_M \delta _M  = \frac{m_1  + m_2}
{m_M}.
\label{GM}
\end{eqnarray}

The branching ratios (BRs)
\begin{equation}
B_{\ell \ell '} =\Gamma (M^ +   \to M'^ -  \ell ^ +  \ell '^ +  )/
\Gamma (M^ +   \to {\rm all})
\label{BR}
\end{equation}
have been calculated for two limiting cases of {\it heavy} ($m_N  \gg m_M $) and {\it light} ($m_N  \ll m_\ell,~m_{\ell '}$) Majorana neutrinos (see Eqs. (\ref{lh}), (\ref{Li})  and Ref. \cite{ABS1} for details):
\begin{equation}
B^{\rm heavy}_{\ell \ell '} = C^{\rm heavy}_{\ell \ell '}\left|{\left\langle {m_{\ell \ell '}^{ - 1} } \right\rangle}\right| ^2 ,~
B^{\rm light}_{\ell \ell '} = C^{\rm light}_{\ell \ell '} \left| { \left\langle {m_{\ell \ell '} } \right\rangle } \right|^2,
\label{BRhl}
\end{equation}
where the {\it effective} Majorana masses are defined as follows:
\begin{equation}
\left\langle {m_{\ell \ell '}^{ - 1} } \right\rangle  =  {\sum\nolimits_N {U_{\ell N} U_{\ell 'N} \eta _N
m^{-1}_N}} ,~
\left\langle {m_{\ell \ell '} } \right\rangle  = {\sum\nolimits_N {U_{\ell N} \,U_{\ell 'N} \eta _N m_N } }.
\label{eff}
\end{equation}
Here the coefficients $C^{\rm heavy}_{\ell \ell '}$ are expressed model independently through the meson decay constants, and $C^{\rm light}_{\ell \ell '}$'s are calculated with use of the model function (\ref{GM}). The following values of the parameters have been used in numerical calculations:
$(f_\pi, f_K, f_D) = (130.7, 159.8, 228)~\mbox{MeV}$; $(m_u, m_d, m_s, m_c) = (4, 7, 150, 1.26\times 10^3)~\mbox{MeV}$. The results are shown in the third and fifth columns of Table~1. The second column of this table shows the present direct experimental upper bounds on the BRs \cite{PDG-08} which are too weak to set reasonable limits on the effective Majorana masses (\ref{eff}).
\begin{table}[h!]
\begin{center}
\begin{tabular}{|c|c|c|c|c|c|}
\hline Rare decay& Exp. upper &$C^{\rm heavy}_{\ell \ell '}$&  Ind.
bound&$C^{\rm light}_{\ell \ell '}$&  Ind.
bound
\\ &{bound} on ${B}_{_{\ell \ell ^{\prime }}}$& $({\rm MeV}^{2})$& on ${
B}^{\rm heavy}_{_{\ell \ell ^{\prime }}} $& $({\rm MeV}^{- 2})$& on ${
B}^{\rm light}_{_{\ell \ell ^{\prime }}} $\\ \hline\hline

$K^{+}\to \pi ^{-}e^{+}e^{+}$ & $6.4\times 10^{-10}$ & $8.5\times
10^{- 10}$ &$5.9\times 10^{-32}$&$4.4\times 10^{-20}$&$2.3\times
10^{-33}$ \\ \hline

$K^{+}\to \pi ^{-}\mu ^{+}\mu ^{+}$ & $3.0\times 10^{-9}$ &
$2.4\times 10^{- 10}$&$1.1\times 10^{-24}$&$1.2\times
10^{-20}$&$6.2\times 10^{- 34}$ \\ \hline

$K^{+}\to \pi ^{-}e^{+}\mu ^{+}$ & $5.0\times 10^{-10}$ &
$1.0\times 10^{- 9}$ &$5.1\times 10^{-24}$&$8.8\times
10^{-20}$&$2.0\times 10^{-33}$ \\ \hline\hline
$D^{+}\to K ^{-}e^{+}e^{+}$ & $4.5\times 10^{-6}$ & $2.2\times
10^{-9}$ &$1.5\times 10^{-31}$&$4.5\times 10^{-21}$&$2.4\times
10^{-34}$ \\ \hline

$D^{+}\to K ^{-}\mu ^{+}\mu ^{+}$ & $1.3\times 10^{-5}$ &
$2.0\times 10^{-9}$ &$8.9\times 10^{- 24}$&$4.1\times
10^{-21}$&$2.2\times 10^{-34}$ \\ \hline

$D^{+}\to K ^{-}e^{+}\mu ^{+}$ & $1.3\times 10^{- 4}$ & $4.2\times
10^{-9}$ &$2.1\times 10^{-23}$&$9.1\times 10^{-21}$&$2.0\times
10^{-34}$ \\ \hline
\end{tabular}
\caption{Experimental and indirect upper bounds on the branching ratios $B_{\ell \ell '} $ for the rare meson decays with $\Delta L = 2$
mediated by heavy or light Majorana neutrinos.}
\label{Tab1}
\end{center}
\vspace{-0.4cm}
\end{table}
So we have derived the {\it indirect} bounds on the BRs (\ref{BRhl}) using the limits on the masses (\ref{eff}) obtained from the precision electroweak measurements, neutrino oscillation experiments, searches for the neutrinoless double beta decay and cosmological data:
\begin{eqnarray}
&\mbox{}\!\left|\left\langle {m_{ee}^{ - 1} } \right\rangle\right| < \left(1.2 \times 10^8~{\rm GeV} \right)^{- 1}\!,\left|\left\langle {m_{\mu \mu }^{ - 1} } \right\rangle\right|  < (1.5 \times 10^4~{{\rm GeV)}}^{ - 1}\!,\left|\left\langle m_{e\mu}^{ - 1} \right\rangle\right|  < (1.4 \times 10^4 \;{{\rm GeV)}}^{ - 1}\!;\nonumber\\
&\left|\left\langle {m_{\ell \ell } } \right\rangle\right|  < 0.23~{\rm eV}~(\ell  = e,~\mu ),~ \left|\left\langle {m_{e\mu } } \right\rangle\right|  < 0.15~{\rm eV}.
\label{lim}
\end{eqnarray}
These indirect bounds (see the forth and sixth columns of Table 1) are greatly more stringent than the direct
ones.

\subsection{Decays in the MSSM with explicit $R$-parity violation}

Here we consider another mechanism of the  $\Delta L =2$ rare
decays (\ref{dec}) based on  $R$-parity violating supersymmetry
(SUSY) (for a review, see Ref.~\cite{Bar}).  The minimal supersymmetric extension of the SM (MSSM) includes the fields of the two-Higgs-doublet extension of the SM and those of the corresponding {\it supersymmetric partners}.  Each fermion (boson) has a superpartner of spin 0 (1/2). $R$-parity
is defined as $R=(-1)^{3(B-L)+2S}$, where $S$, $L$, and $B$ are the spin, the lepton and baryon numbers, respectively. In the MSSM,
$R$-parity conservation is imposed to prevent the $L$ and $B$
violation; it also leads to  the production of superpartners in pairs and
ensures the stability of the lightest superparticle. However, {\it neither gauge
invariance nor supersymmetry require $R$-parity conservation}.
There are many generalizations of the MSSM with explicitly or
spontaneously broken $R$-symmetry \cite{Bar}. We consider a SUSY
theory with the minimal particle content of the MSSM and explicit
$R$-parity violation ($\not\!\! R$MSSM).

The most general form for the $R$-parity and lepton number
violating part of the superpotential is given by
\vspace{-0.2cm}
\begin{equation}
\label{rpv}
W_{\not R}  = \varepsilon _{\alpha\beta } \left(
{\frac{1} {2}\lambda _{ijk} L_i^\alpha  L_j^\beta  \bar E_k  +
\lambda '_{ijk} L_i^\alpha  Q_j^\beta  \bar D_k  + \epsilon _i
L_i^\alpha H_u^\beta  } \right).
\vspace{-0.2cm}
\end{equation}
Here $i, j, k =1, 2, 3$ are generation indices, $L$ and $Q$ are
$SU(2)$ doublets of left-handed lepton and quark superfields
($\alpha, \beta = 1, 2$ are isospinor indices), $\bar{E}$ and
$\bar{D}$ are singlets of right-handed superfields of leptons and
down quarks, respectively; $H_u$ is a doublet Higgs superfield
(with hypercharge $Y=1$); $\lambda _{ijk} = - \lambda _{jik}
,~\lambda '_{ijk}$ and $\epsilon _i $ are constants.

In the superpotential (\ref{rpv}) the trilinear ($\propto \lambda
,~\lambda '$) and bilinear ($\propto \epsilon $) terms are present.
At first, we assume that the bilinear terms are absent at tree
level ($\epsilon =0$). They will be generated by quantum
corrections~\cite{Bar}, but it is expected that the phenomenology
will still be dominated by the tree-level trilinear terms.

The leading order amplitude of the process $ K^{+}\to \pi^{-}+
\ell^{+} + \ell^{'+}$ in the $\not\!\! R$MSSM is described by three types
of diagrams shown in Fig. 2.
\begin{figure}[h!]
\vspace{-0.2cm}
\centering
\includegraphics[scale=0.7]{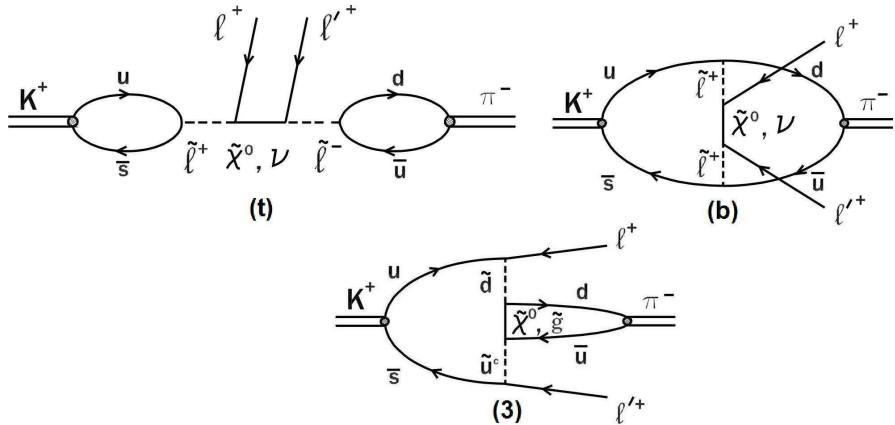}
\vspace{-1.2cm}
\caption{Feynman diagrams for the  decay  $
K^{+}\to \pi^{-}+ \ell^{+}+ \ell^{'+}$ mediated by Majorana
neutrinos $\nu$, neutralinos $\tilde \chi ^0 $, gluinos $\tilde g$
with $\tilde f$ being the scalar superpartners of the
corresponding fermions $f=\ell, u, d$ (leptons and quarks).}
\label{Fig:decR}
\end{figure}
For the numerical estimates of the branching ratios (\ref{BR}), we have used the known values for
the couplings, decay constants, meson, lepton and current quark
masses \cite{PDG-08}, and a typical set of the MSSM parameters and the elements of the $4\times 4$ neutralino mixing matrix
from Ref. \cite{ADD}. In addition, we have taken all the masses of superpartners to be equal with a common value
$m_{SUSY}$. Taking into account the present bounds on the
effective Majorana masses (\ref{lim}), we find that the
main contribution to the decay width  comes from the exchange by
neutralinos and gluinos (see Fig.~2). The results of the
calculations for the decays  $K^+ \to \pi^- \ell^+\ell^+$ and $D^+ \to K^- \ell ^+ \ell '^+$ are shown in Table 2 (here $ m_{200}  = m_{SUSY} /(200~
\mbox{GeV})$) \cite{ABS2}.

\begin{wraptable}{r}{0.64\textwidth}
\vspace{-0.8cm}
\begin{center}
\begin{tabular}{|c|c|}
\hline Rare decay&$B_{\ell
\ell^{\prime }}\times m_{200}^{10}$\\ \hline\hline $K^{+}\to \pi
^{-}e^{+}e^{+}$&$1.3\times
10^{-17}|\lambda'_{111}\lambda'_{112}|^2$
\\ \hline

$K^{+}\to \pi ^{-}\mu ^{+}\mu ^{+}$&$4.7\times 10^{-18}|\lambda'_{211}\lambda'_{212}|^2$
\\ \hline

$K^{+}\to \pi ^{-}e^{+}\mu ^{+}$ & $4.3\times
10^{-18}|\lambda'_{111}\lambda'_{212}+\lambda'_{211}\lambda'_{112}|^2$
\\ \hline\hline
$D^{+}\to K ^{-}e^{+}e^{+}$ &$1.4\times
10^{-18}|\lambda'_{122}\lambda'_{111}-0.39\lambda'_{121}\lambda'_{112}|^2$
\\ \hline

$D^{+}\to K ^{-}\mu ^{+}\mu ^{+}$ &$1.3\times
10^{-18}|\lambda'_{222}\lambda'_{211}-0.39\lambda'_{221}\lambda'_{212}|^2$
\\ \hline
$D^{+}\to K ^{-}e^{+}\mu ^{+}$ &$6.5\times
10^{-19}|(\lambda'_{122}\lambda'_{211}+\lambda'_{222}\lambda'_{111})$\\
&$-0.39(\lambda'_{121}\lambda'_{212}+\lambda'_{221}\lambda'_{112})|^2$
\\ \hline
\end{tabular}
\caption{The branching ratios $B_{_{\ell \ell ^{\prime }}}$
for the meson decays mediated by trilinear Yukawa couplings in the $\not\!\! R$MSSM.}
\end{center}
\label{Tab2}
\vspace{-0.5cm}
\end{wraptable}
For upper bounds on the trilinear couplings (from analysis of a number of other processes~\cite{Bar}) $\left| {\lambda '\lambda '} \right| \lesssim 5\times 10^{-6}$, we obtain an estimate of the BRs:
\begin{equation}
B_{\ell \ell^{\prime }}({\rm tri}\!{\not\!R})\lesssim 10^{-28}m_{200}^{-10}.~
\label{tri}
\end{equation}
This estimate is much smaller than the corresponding direct
experimental bounds and lies between (except for the $ee$ decay mode)
the indirect bounds based on the mechanisms of the decays mediated by heavy and
light Majorana neutrinos (see Table~1).

For the case of tree-level bilinear couplings ($\epsilon \neq 0, \lambda = 0,  \lambda' = 0$ in Eq. (\ref{rpv})), trilinear couplings cannot be generated via radiative corrections \cite{Bar}. The bilinear terms in the superpotential induce mixing between the SM leptons and the MSSM charginos and neutralinos in the mass-eigenstate basis and lead to the $\Delta L = \pm 1$ lepton-quark interactions, in particular, giving rise  to the meson decays (\ref{dec}). For this bilinear decay mechanism, the order-of-magnitude estimate of the BRs  is given by \cite{ABS3}
\begin{equation}
B_{\ell \ell^{\prime }}({\rm bi}\!{\not\!R})\lesssim 10^{-48}m_{200}^{-10},
\label{bi}
\end{equation}
which is twenty orders of magnitude smaller than that for the trilinear mechanism (\ref{tri}).

We note that our results for the decay $K^+ \to \pi^- \mu^+\mu^+$ (see Tables 1 and 2) are in agreement with the corresponding estimates obtained in Refs. \cite{LS1,LS2}.

\section{Conclusion}

In the minimal SM (with massless neutrinos), each lepton family number $L_\ell$ and the baryon number $B$ are conserved  due to the
accidental global symmetry (\ref{GSMg}). 

The unambiguous observation of neutrino oscillations implies nonzero neutrino masses and lepton mixing and clearly demonstrates the LFN violation (with conservation of the total LN).

It is natural to believe that the neutrino mass is the first evidence of New Physics.

The LN violation is a generic feature of theories beyond the SM, and  searching $\Delta L\neq 0$ processes is a way to test these theories.

The semileptonic rare meson decays (RMDs) with $\Delta L = 2$ were investigated in the SM extended by Majorana neutrinos and in the MSSM with explicit $R$-parity violation. The indirect bounds  on the RMD branching ratios have been derived from the precision electroweak measurements, neutrino oscillation experiments, searches for the $0\nu2\beta$ decay, cosmological data, and bounds on $R$-parity violating couplings. These indirect bounds are {\it greatly more stringent} than the bounds from direct searching RMDs. So the RMDs will hardly be seen in the nearest future.

The neutrinoless double beta decay and the production of same-sign dileptons at colliders (like the LHC) look substantially more promising for observation.

\section*{Acknowledgments}

I am grateful to Ahmed Ali, Nikolai Zamorin, Maria Sidorova and Dmitri Zhuridov for fruitful collaboration and to Nikolai Nikitin for useful discussions, as well as to the organizers  of the Helmholtz International Summer School ``Heavy Quark Physics" (Dubna, August 2008), Ahmed Ali and Mikhail Ivanov, for inviting me to give this lecture and creating a nice working atmosphere.


\begin{footnotesize}



%

\end{footnotesize}


\end{document}